\documentclass[12pt]{article}
\usepackage{epsfig}
\begin{document}


\title{\flushright{\small MIT-CTP \# 3669}\\ 
\center{Conditions for existance of neutral strange quark matter}
\footnote{This work was completed in November, 2003}}

\author{Elena Gubankova\\
{\it Center for Theoretical Physics, Department of Physics},\\
{\it Massachusetts Institute of Technology, Cambridge, MA 02139}}

\maketitle

\begin{abstract}

Breached pairing
solutions to the gap equation are obtained analytically 
in for two and three quarks and for low and high temperatures. 
We compare the energy of these states to that of other homogeneous states
under the condition of electric neutrality.
We found the two-flavor
BP and the three flavor mixed BCS-BP phases, which are stable over a wide range of
parameters. Both phases contain four BP modes in the quasiparticle spectrum.

\end{abstract}

\section{Introduction}

Recently there has been considerable interest in a class of possible new states of matter
featuring coexistence of superfluid and normal components.  
Examples have appeared in several
variants, under various names (``Sarma state'' \cite{Sarma}, ``interior gap'' \cite{LiuWilczek}, 
``breached pair''\cite{GubankovaLiuWilczek} - which we adopt here, ``gapless superconductor'' \cite{ShovkovyHuang}).
Breached pair states are candidates to arise when there are interactions favoring pairing between fermions 
with fermi surfaces of different size.  They are quite different from
the famous LOFF phases \cite{LOFF}, and need not break translation invariance.   
The basic {\it ansatz\/} for this class of states goes back to old work of Sarma \cite{Sarma}.  
There are two main causes for the recent 
upsurge in interest. First, new candidate applications have emerged.  
These include notably cold atom systems, where there can be great flexibility in manipulating 
denisities, effective masses, and interactions \cite{experiment}; and high-density QCD, relevant to neutron star interiors, 
as will be our concern below.  Second, and importantly, 
parameter regimes have been identified where the new states are likely to be stable.  
General heurisitic arguments, based on extrapolation from clear limiting cases (ultrastrong coupling, flat bands) 
were presented in \cite{LiuWilczek}. Quantitative comparisons
between breached pairing and other homogeneous states, identifying favorable cases, 
have been presented in \cite{GubankovaLiuWilczek}.   

It has also been suggested \cite{Bedaque} that inhomogeneous 
(phase separated superfluid and normal) states can compete favorably with breached pair states.  
This is an interesting question, that receives further investigation. Our preliminary conclusion is that such phase 
separation can occur, but only in rather special 
parameter regimes. We shall not discuss it further here.
  
In this paper we shall compare the energy of breached pair to conventional BCS and normal state alternatives, 
in some models suggested by high density QCD.  Our
calculations may have application to neutron stars, but in view of the difficulty of interpreting 
astrophysical observations and the idealized nature of the models 
perhaps their main interest is methodological.  
We illustrate in this specific context how one establishes the (in)stability of breached pairing, 
and the influence of different physical conditions.

\section{Breached pairing superconductor at zero temperature}

Here we summarize properties of the breached pairing superconductor 
concetrating on the stability condition of the breached pairing phase.
We present calculations in a toy model \cite{GubankovaLiuWilczek} that has a superfluid ground state 
of up-strange quark pairs. The relation of more realistic cases to this model 
is worked out in the Appendix.  

We consider massive $s$ and massless $u$ quarks both being relativistic,
\begin{equation}
\varepsilon_{{\bf p}}^{u} = p-p_{F}^{u} \,, \qquad 
\varepsilon_{{\bf p}}^{s} =p-p_{F}^{s} \label{eq:1.1}
\end{equation} 
where the fermi momenta are related to the chemical potentials as
$p_{F}^{u} =\mu_{u}$ and $p_{F}^{s} =\sqrt{\mu_{s}^{2}-m_{s}^2}$, and the fermi momentum
for the $s$-quark is smaller than that for the $u$-quark, $p_{F}^{s}<p_{F}^{u}$.
We suppress color indices and postulate a
a weak attractive interaction
$g(\psi_{u}\sigma_{2}\psi_{s})(\psi^{\dagger}_{u}\sigma_{2}\psi^{\dagger}_{s})$ 
between the light and heavy species.
In a basis of light particles and heavy holes the quadratic
part of the action is
\begin{eqnarray}
&& S=\sum_{\bf p}
\left(\begin{array}{cc}
\psi^{\dagger}_{u{\bf p}}&\psi_{s-{\bf p}}
\end{array}\right)
\left(\begin{array}{cc}
-\varepsilon_{{\bf p}}^{u}&\Delta\\
\Delta^{\ast}&\varepsilon_{{\bf p}}^{s}\end{array}\right)
\left(\begin{array}{c}
\psi_{u{\bf p}}\\\psi^{\dagger}_{s-{\bf p}}
\end{array}\right)
\label{eq:0.1}\end{eqnarray} 
where the gap parameter is defined as 
$\Delta^{\ast}=g/V\sum_{\bf p}
\langle \psi^{\dagger}_{u{\bf p}}\psi^{\dagger}_{s-{\bf p}}\rangle_{BP}$
in a breached pairing (BP) superconducting ground state, and
momentum sum is $\sum_{\bf p}=V\int d^{3}p/(2\pi)^3$.
Since the gap parameter $\Delta$ is defined as a $c$-number, this action
can be diagonalized. The quasiparticle energies are
$\delta p_{F}\pm\sqrt{\varepsilon_{\bf p}^2+\Delta^2}$, where
$\varepsilon_{\bf p}=(\varepsilon_{\bf p}^{u}+\varepsilon_{\bf p}^{s})/2
=p-p_{F}$ with the average Fermi momentum $p_{F}=(p_{F}^{u}+p_{F}^{s})/2$
and the mismatch in Fermi momenta 
$\delta p_{F}=(\varepsilon_{\bf p}^{s}-\varepsilon_{\bf p}^{u})/2
=(p_{F}^{u}-p_{F}^{s})/2>0$. 
We obtain the following free energy density \cite{GubankovaLiuWilczek}
\begin{equation}
\Omega_{BP}=\frac{\Delta^2}{g}+\int\frac{d^{3}p}{(2\pi)^3}\left(
\varepsilon_{\bf p}-\sqrt{\varepsilon_{\bf p}^2+\Delta^2}
\right)-\int_{R}
\frac{d^{3}p}{(2\pi)^3}\left(
\delta p_{F}-\sqrt{\varepsilon_{\bf p}^2+\Delta^2}\right)
\label{eq:0.2}\end{equation}
where there is no pairing in the momentum range 
$R=\{|\varepsilon_{\bf p}|\leq\sqrt{\delta p_{F}^2-\Delta^2}\}$,
which is singly occupied by $u$-quarks.  
The gap equation is found either by minimizing the free energy density
or as a self-consistent condition on $\Delta$
\begin{equation}
\frac{2\Delta}{g}=\int\frac{d^{3}p}{(2\pi)^3}
\frac{\Delta}{\sqrt{\varepsilon_{\bf p}^2+\Delta^2}}
-\int_{R}\frac{d^{3}p}{(2\pi)^3}
\frac{\Delta}{\sqrt{\varepsilon_{\bf p}^2+\Delta^2}}
\label{eq:0.3}\end{equation}
Writing $\int\frac{d^{3}p}{(2\pi)^3}=N(0)\int_{-\omega}^{\omega}d\varepsilon_{\bf p}$,
where the density of states is
$N(0)=\int\frac{d^{3}p}{(2\pi)^3}\delta(\varepsilon_{\bf p})$ 
and the UV cutoff is $\omega\sim p_{F}$, we have
\begin{equation}
\frac{1}{gN(0)}=\ln\left(\frac{2\omega}{\Delta}\right)
-\ln\left(\frac{\delta p_{F}+\sqrt{\delta p_{F}^2-\Delta^2}}{\Delta}\right)
\label{eq:0.3a}\end{equation}
Introducing the BCS gap 
at zero Fermi momentum mismatch, $\Delta_{0}$, 
we find the breached pairing and the BCS solutions 
\begin{eqnarray}
&&\hspace{-3cm} \Delta=\left[\Delta_{0}(2\delta p_{F}-\Delta_{0})\right]^{1/2}\qquad
(\delta p_{F}>\Delta)\nonumber\\
&&\hspace{-3cm} \Delta = \Delta_{0} \qquad (\delta p_{F}\leq \Delta)
\label{eq:0.3b}\end{eqnarray}
 
Stability of the breached pairing state depends on the physical conditions
imposed on the system \cite{GubankovaLiuWilczek}. Solving these constraints
leads to the dependence of the Fermi momentum mismatch after pairing 
on the gap parameter $\delta p_{F}(\Delta)$, which is parametrized 
\cite{GubankovaLiuWilczek}
\begin{equation}
\delta p_{F}(\Delta)=\frac{\delta p_{F}}
{1-\alpha^2\frac{2\Delta^2}{\Delta_{0}^2+\Delta^2}}
\label{eq:0.4}\end{equation}
with a constant $\alpha^2$ depending on a condition and $\delta p_{F}$ is the mismatch
before pairing. 
Substituting $\delta p_{F}\rightarrow \delta p_{F}(\Delta)$ 
in the solution of the gap equation, Eq.~(\ref{eq:0.3b}), we obtain
\begin{equation}
\Delta=\left[\Delta_{0}(\Delta_{0}-2\delta p_{F})/(2\alpha^2-1)\right]^{1/2}
\label{eq:0.5}\end{equation}
For $\alpha^2> 1/2$ one has a stable breached pairing solution \cite{GubankovaLiuWilczek}.

One can check stability of the BP state also directly calculating the condensation energy,
which is obtained by integrating the gap equation, Eq.~(\ref{eq:0.3}), over the gap parameter
\cite{RajagopalWilczek}
\begin{equation}
\Omega_{BP}-\Omega_{N}=\int_{0}^{\Delta}d\Delta^{\prime}\left(
-\frac{2\Delta^{\prime}}{g}+\int\frac{d^{3}p}{(2\pi)^3}
\frac{\Delta^{\prime}}{\sqrt{\varepsilon_{\bf p}^2+\Delta^{2}}}
-\int_{R}\frac{d^{3}p}{(2\pi)^3}
\frac{\Delta^{\prime}}{\sqrt{\varepsilon_{\bf p}^2+\Delta^{2}}}\right)
\label{eq:0.6}\end{equation}  
where $\Delta$ is a solution of the gap equation, $\Omega_{N}$ is the energy density
in the normal state.
We use the gap equation to eliminate $g$,
\begin{equation}
\Omega_{BP}-\Omega_{N}= 2N(0)\int_{0}^{\Delta}\Delta^{\prime}d\Delta^{\prime}\left(
\ln\left(\frac{\Delta^{\prime}}{\Delta}\right)
-\ln\left(\frac{\Delta^{\prime}}{\Delta}\right)
+\ln\left(\frac{\delta p_{F}(\Delta^{\prime})
+\sqrt{\delta p_{F}(\Delta^{\prime})^2-\Delta^{\prime\,2}}}{\delta p_{F}(\Delta)
+\sqrt{\delta p_{F}(\Delta)^2-\Delta^{2}}}\right)
\right)\nonumber\\
\label{eq:0.7}\end{equation}
where the first term gives the BCS condensation energy, $-N(0)\Delta^2/2$,
which is cancelled by the second term arising from integrating in the BP region $R$.  
Integrating by parts, we find
\begin{eqnarray}
\Omega_{BP}-\Omega_{N}= - N(0)\int_{0}^{\Delta}
\frac{\Delta^{\prime\,2}d\Delta^{\prime}}{\sqrt{\delta p_{F}(\Delta^{\prime})^{\,2}-\Delta^{\prime\,2}}}
\left[\frac{d\delta p_{F}(\Delta^{\prime})}{d\Delta^{\prime}}
-\frac{\Delta^{\prime}}{\delta p_{F}(\Delta^{\prime})
+\sqrt{\delta p_{F}(\Delta^{\prime})^{\,2}-\Delta^{\prime\,2}}}
\right]\nonumber\\
\label{eq:0.8}\end{eqnarray}   
Using in Eq.~(\ref{eq:0.8}) the parametrization
$\delta p_{F}(\Delta)=\delta p_{F}(1+\alpha^2\frac{\Delta^2}{2\delta p_{F}^2})$, which is
valid for small $\Delta$ and is equivalent to Eq.~(\ref{eq:0.4}), we find 
the condensation energy of the BP state in the leading order $O(\Delta^2/\delta p_{F}^2)$
\begin{equation}
\Omega_{BP}-\Omega_{N}=-N(0)\frac{\Delta^{4}(2\alpha^2-1)}{8\delta p_{F}^2}
\label{eq:0.10}\end{equation}  
that shows a stable state when $\alpha^2>1/2$. In the third section, we calculate
$\alpha^2$ for different BP solutions under condition of the electric neutrality,
checking stability of these solutions.

Differentiating the free energy density, Eq.~(\ref{eq:0.2}), 
with respect to the quark chemical potentials,
$\mu_{u}$ and $\mu_{s}$, we obtain the corresponding quark number densities. 
Namely, 
$n_{u}+n_{s}=\partial\Omega_{BP}/\partial p_{F}$ and
$n_{u}-n_{s}=\partial\Omega_{BP}/\partial \delta p_{F}$,
which are equal to 
\begin{eqnarray}
n_{u}+n_{s}=\int\frac{d^{3}p}{(2\pi)^3}\left(
1-\frac{\varepsilon_{\bf p}}{\sqrt{\varepsilon_{\bf p}^2+\Delta^2}}\right)
+\int_{R}\frac{d^{3}p}{(2\pi)^3}
\frac{\varepsilon_{\bf p}}{\sqrt{\varepsilon_{\bf p}^2+\Delta^2}}~,~
n_{u}-n_{s}=\int_{R}\frac{d^{3}p}{(2\pi)^3}1
\label{eq:0.11}\end{eqnarray}   
Integrating, we obtain
\begin{eqnarray}
n_{u}+n_{s}=2N(0)\,\frac{p_{F}}{3}~,~
n_{u}-n_{s}=2N(0)\sqrt{\delta p_{F}(\Delta)^2-\Delta^2}
\label{eq:0.11a}\end{eqnarray} 
where the density of states is $N(0)=p_{F}^2/\pi^2$.
When $\Delta=0$,  
$n_{u}=p_{F}^2/(3\pi^2)(p_{F}+3\delta p_{F})\approx p_{F}^{u\,3}/(3\pi^2)$
and  $n_{s}=p_{F}^2/(3\pi^2)(p_{F}-3\delta p_{F})\approx p_{F}^{s\,3}/(3\pi^2)$,
since $\delta p_{F}\ll p_{F}$,
and we recover the free gas limit. This means that the correct way to construct
the BP state is to first fill noninteracting quark states up to the corresponding
Fermi momenta and to then pair that produces the condensation energy, $\Omega_{cond}$.
In contrast to that, in the BCS state we first fill free quark states up to the common
Fermi momentum, $p_{F}$, and then pair \cite{RajagopalWilczek}.
Therefore, for the BCS state,
we gain in the condensation energy, 
$-N(0)\Delta_{0}^2/2$, but lose by bringing two Fermi surfaces together, $N(0)\delta p_{F}^2$
\cite{RajagopalWilczek}. However, the condensation energy of the BP state is parametrically
smaller, $\sim \Delta^4$ Eq.~(\ref{eq:0.10}), than that of the BCS state.

\section{Breached pairing at finite temperature} 

\subsection{Gap equation and quark densities}

We generalize our model to the case of nonzero temperature.
The partition function at finite $T=1/\beta$ is 
\begin{equation}
Z = {\rm Tr}\,{\rm e}^{H-\mu_{u} n_{u}-\mu_{s}n_{s}}=
i\int D\psi^{\dagger}_{\alpha,n}({\bf p})
D\psi_{\alpha,n}({\bf p}){\rm e}^{S} 
\label{eq:1.2}
\end{equation}
where $n_{u}, n_{s}$ are the number of $u$ and $s$-quarks, and functional integral
is performed in momentum-frequency and  
in flavor spaces, $\prod_{n}\prod_{\bf p}\prod_{\alpha}
d\psi^{\dagger}_{\alpha,n}({\bf p})d\psi_{\alpha,n}({\bf p})$.
Assuming a superconducting {\it ansatz\/} for the ground state, 
in a basis of light particles and heavy holes 
the quadratic part of the action takes the form
\begin{eqnarray}
&& S=\sum_{n}\sum_{\bf p}i
\left(\begin{array}{cc}
\psi^{\dagger}_{u,n}({\bf p})&\psi_{s,n}(-{\bf p})
\end{array}\right)
K_{\alpha\beta}
\left(\begin{array}{c}
\psi_{u,n}({\bf p})\\\psi^{\dagger}_{s,n}(-{\bf p})
\end{array}\right)\nonumber\\
&& K_{\alpha\beta}=
(i\beta)
\left(\begin{array}{cc}
i\omega_{n}-\varepsilon_{{\bf p}}^{u}&\Delta\\
\Delta^{\ast}&i\omega_{n}+\varepsilon_{{\bf p}}^{s}\end{array}\right)
\label{eq:1.3}\end{eqnarray}
where the gap parameter is defined as
$\Delta^{\ast}=g/V\sum_{n}\sum_{\bf p}
\langle \psi^{\dagger}_{u,n}({\bf p})\psi^{\dagger}_{s,n}(-{\bf p})\rangle_{BP}$.
The momentum sum is $\sum_{\bf p}=V\int d^{3}p/(2\pi)^3$ and the frequency sum involves
$\omega_{n}=(2n+1)\pi T$. Integrating over Grassmann variables, we have 
$Z={\rm det}K$, where the determinant is carried out over the flavor indices
and in momentum-frequency space. Employing the identity $\ln {\rm det} K ={\rm Tr}\ln K$,
we find 
\begin{equation}
\ln Z=\sum_{n}\sum_{\bf p}\ln\left[\beta^2\left((\omega_{n}-i\delta\varepsilon_{\bf p})^2+
\varepsilon_{\bf p}^2+\Delta^2\right)\right]
\label{eq:1.4}\end{equation}
where $\delta\varepsilon_{\bf p}=(\varepsilon_{\bf p}^{s}-\varepsilon_{\bf p}^{u})/2$ 
and
$\varepsilon_{\bf p}=(\varepsilon_{\bf p}^{s}+\varepsilon_{\bf p}^{u})/2$. 
Since both positive and negative frequencies
are summed over, Eq.~(\ref{eq:1.4}) can be rewritten
\begin{equation}
\ln Z=\frac{1}{2}\sum_{n}\sum_{\bf p}\left(\phantom{\frac{1}{1}}\hspace{-0.3cm}
\ln\left[\beta^2\left(\omega_{n}^2+
(\sqrt{\varepsilon_{\bf p}^2+\Delta^2}+\delta\varepsilon_{\bf p})^2\right)\right]
+\ln\left[\beta^2\left(\omega_{n}^2+
(\sqrt{\varepsilon_{\bf p}^2+\Delta^2}-\delta\varepsilon_{\bf p})^2\right)\right]\right)
\label{eq:1.5}\end{equation}

Following \cite{Kapusta} we write
\begin{equation}
\ln\left[(2n+1)^2\pi^2+\beta^2(\omega_{\bf p}\pm\delta\varepsilon_{\bf p})^2\right]
=\int_{1}^{\beta^2(\omega_{\bf p}\pm\delta\varepsilon_{\bf p})^2}\frac{d\theta^2}{\theta^2+(2n+1)^2\pi^2}
+\ln\left[1+(2n+1)^2\pi^2\right]
\label{eq:1.6}\end{equation}
where $\omega_{\bf p}=\sqrt{\varepsilon_{\bf p}^2+\Delta^2}$. The sum over $n$ may be carried out
by using the summation formula \cite{Kapusta}, \cite{GradshteynRyzhik}
\begin{equation}
\sum_{n=-\infty}^{\infty}\frac{1}{(2n+1)^2\pi^2+\theta^2}
=\frac{1}{\theta}\left(\frac{1}{2}-\frac{1}{{\rm e}^{\theta}+1}\right)
\label{eq:1.7}\end{equation}
Integrating over $\theta$, and dropping terms independent of $\beta$ and $\delta\varepsilon_{\bf p}$,
we finally obtain
\begin{eqnarray}
\ln Z &=& \frac{1}{2}V\int\frac{d^{3}p}{(2\pi)^3}
\left\{\beta(\sqrt{\varepsilon_{\bf p}^2+\Delta^2}-\delta\varepsilon_{\bf p})
+\beta(\sqrt{\varepsilon_{\bf p}^2+\Delta^2}+\delta\varepsilon_{\bf p})\right. \nonumber\\
&+&\left. 2\ln\left[1+{\rm e}^{-\beta(\sqrt{\varepsilon_{\bf p}^2+\Delta^2}-\delta\varepsilon_{\bf p})}\right]
+2\ln\left[1+{\rm e}^{-\beta(\sqrt{\varepsilon_{\bf p}^2+\Delta^2}+\delta\varepsilon_{\bf p})}\right]\right\}
\label{eq:1.8}\end{eqnarray}

The thermodynamic potential is defined as  
$\Omega=-P=-T\ln Z/V$. 
Adding the contribution from the mean field potential,
$\Delta^2/g$, and
the vacuum energy arising from the normal ordering, 
$\varepsilon_{\bf p}$, 
we obtain
\begin{eqnarray}
\Omega_{BP} &=& \frac{\Delta^2}{g}+\int \frac{d^{3}p}{(2\pi)^3}\left\{\varepsilon_{\bf p}
-\sqrt{\varepsilon_{\bf p}^2+\Delta^2}\right.\nonumber\\
&-&\left.T \ln\left[1+{\rm e}^{-\beta(\sqrt{\varepsilon_{\bf p}^2+\Delta^2}-\delta p_{F})}\right]
-T\ln\left[1+{\rm e}^{-\beta(\sqrt{\varepsilon_{\bf p}^2+\Delta^2}+\delta p_{F})}\right]\right\}
\label{eq:1.10}\end{eqnarray}
where $\varepsilon_{\bf p}=p-p_{F}$, 
$\delta\varepsilon_{\bf p}=\delta p_{F}$, with
$p_{F}=(p_{F}^{u}+p_{F}^{s})/2$ and $\delta p_{F}=(p_{F}^{u}-p_{F}^{s})/2>0$.
The thermodynamic potential in the four-dimensional notations,
Eq.~(\ref{eq:1.4}), is
\begin{equation}
\Omega_{BP} =\frac{\Delta^2}{g}+T\sum_{n} \int \frac{d^{3}p}{(2\pi)^3}\left\{\varepsilon_{\bf p}
-\ln\left[\beta^2\left((\omega_{n}-i\delta p_{F})^2+
\varepsilon_{\bf p}^2+\Delta^2\right)\right]\right\}
\label{eq:1.11}\end{equation}
where $T\sum_{n}1=T\int_{0}^{\beta}d\tau=1$.
Variation of the thermodynamic potential with respect to the gap parameter,
$\partial\Omega_{BP}/\partial\Delta =0$, gives the gap equation in the $3$-d notations
\begin{eqnarray}
\frac{2\Delta}{g}=\int\frac{d^{3}p}{(2\pi)^3}\frac{\Delta}{\sqrt{\varepsilon_{\bf p}^2+\Delta^2}}
\left[1-\frac{1}{{\rm e}^{\beta(\sqrt{\varepsilon_{\bf p}^2+\Delta^2}-\delta p_{F})}+1}
-\frac{1}{{\rm e}^{\beta(\sqrt{\varepsilon_{\bf p}^2+\Delta^2}+\delta p_{F})}+1}\right]
\label{eq:1.12}\end{eqnarray}
rewritten as
\begin{eqnarray}
\frac{2\Delta}{g}=\int\frac{d^{3}p}{(2\pi)^3}\frac{\Delta}{\sqrt{\varepsilon_{\bf p}^2+\Delta^2}}
\,\frac{1}{2}\left[\tanh\left(\frac{\beta(\sqrt{\varepsilon_{\bf p}^2+\Delta^2}-\delta p_{F})}{2}\right)
+\tanh\left(\frac{\beta(\sqrt{\varepsilon_{\bf p}^2+\Delta^2}+\delta p_{F})}{2}\right)\right]
\nonumber\\
\label{eq:1.12a}\end{eqnarray}
and the gap equation in the $4$-d notations
\begin{eqnarray}
\frac{2\Delta}{g}=T\sum_{n}\int \frac{d^{3}p}{(2\pi)^3}\frac{\Delta}
{(\omega_{n}-i\delta p_{F})^2+\varepsilon_{\bf p}^2+\Delta^2}
\label{eq:1.13}\end{eqnarray}
where $\int\frac{d^{3}p}{(2\pi)^3}=N(0)\int_{-\omega}^{\omega}d\varepsilon_{\bf p}$,
with the density of states 
$N(0)=\int\frac{d^{3}p}{(2\pi)^3}\delta(\varepsilon_{\bf p})$ 
and the UV cutoff $\omega\sim p_{F}$.
 
The poles of the anomalous propagator are located at
$p_0=i\omega_{n}=\pm\sqrt{\varepsilon_{p}^2+\Delta^2}-\delta p_{F}$, which define
the quasiparticle energies. They match energies at $T=0$.

Differentiating the thermodynamic potential with respect to the quark chemical potentials,
$n_{u}+n_{s}=\partial\Omega_{BP}/\partial p_{F}$ and
$n_{u}-n_{s}=\partial\Omega_{BP}/\partial \delta p_{F}$,
we obtain the quark number densities in the $3$-d
\begin{eqnarray}
n_{u}+n_{s} &=& \int\frac{d^{3}p}{(2\pi)^3}
\left[1-\frac{\varepsilon_{\bf p}}{\sqrt{\varepsilon_{\bf p}^2+\Delta^2}}
\left(1-\frac{1}{{\rm e}^{\beta(\sqrt{\varepsilon_{\bf p}^2+\Delta^2}-\delta p_{F})}+1}
-\frac{1}{{\rm e}^{\beta(\sqrt{\varepsilon_{\bf p}^2+\Delta^2}+\delta p_{F})}+1}
\right)\right]
\nonumber\\
n_{u}-n_{s} &=& \int\frac{d^{3}p}{(2\pi)^3}\left[
\frac{1}{{\rm e}^{\beta(\sqrt{\varepsilon_{\bf p}^2+\Delta^2}-\delta p_{F})}+1}
-\frac{1}{{\rm e}^{\beta(\sqrt{\varepsilon_{\bf p}^2+\Delta^2}+\delta p_{F})}+1}\right]
\label{eq:1.14}\end{eqnarray}
and in the $4$-d,
\begin{eqnarray}
n_{u}+n_{s} &=& 2T\sum_{n}\int\frac{d^{3}p}{(2\pi)^3}\left[\frac{1}{2}
-\frac{\varepsilon_{\bf p}}{(\omega_{n}-i\delta p_{F})^2+\varepsilon_{\bf p}^2+\Delta^2}\right]
\nonumber\\
n_{u}-n_{s}&=& 2T\sum_{n}\int\frac{d^{3}p}{(2\pi)^3}
\frac{i(\omega_{n}-i\delta p_{F})}{(\omega_{n}-i\delta p_{F})^2+\varepsilon_{\bf p}^2+\Delta^2}
\label{eq:1.15}\end{eqnarray}

Both the $3$-d representations Eq.~(\ref{eq:1.12},\ref{eq:1.14}) and the $4$-d
Eq.~(\ref{eq:1.13},\ref{eq:1.15}) representations will be useful later.

\subsection{Solving the gap equation}

{\it Small temperatures, $T\ll T_{c}$.}

We now examine how the size of the gap in the energy spectrum depends on the mismatch
of the fermi momenta and on the temperature. 

First consider the case of low temperatures,
$T\ll T_{c}$ where $T_{c}$ is the critical temperature for the BCS superconductor,
and make a suitable expansion of the gap equation, Eq.~(\ref{eq:1.12}).
We have
\begin{equation}
1=gN(0)\int_{0}^{\omega} \frac{d\varepsilon}{\sqrt{\varepsilon^2+\Delta^2}}
\left[1-\frac{1}{{\rm e}^{(\sqrt{\varepsilon^2+\Delta^2}+\delta p_{F})/T}+1}
-\frac{1}{{\rm e}^{(\sqrt{\varepsilon^2+\Delta^2}-\delta p_{F})/T}+1}\right]
\label{eq:1.18}\end{equation}
where $N(0)= p_{F}^2/(2\pi^2)$ is the density of states, 
$\varepsilon=p-p_{F}$, and $\omega$ is the ultraviolet cutoff.
Introducing a dimensionless variable $x=\sqrt{\varepsilon^2+\Delta^2}/\Delta$,
we rewrite the gap equation as 
\begin{equation}
\frac{1}{gN(0)}=\int_{1}^{\omega/\Delta} 
\frac{dx}{\sqrt{x^2-1}}
\left[1-\frac{1}{{\rm e}^{(\Delta/T)(x+\delta p_{F}/\Delta)}+1}
-\frac{1}{{\rm e}^{(\Delta/T)(x-\delta p_{F}/\Delta)}+1}
\right]
\label{eq:1.19}
\end{equation}  
In the limit $\Delta/T\gg 1$ and when $\delta p_{F}>\Delta$, we have
\begin{eqnarray}
\frac{1}{gN(0)}=\int_{1}^{\omega/\Delta} 
\frac{dx}{\sqrt{x^2-1}}
&-&\int_{1}^{\infty}\frac{dx}{\sqrt{x^2-1}}
\sum_{n=1}^{\infty}(-1)^{n+1}{\rm e}^{-n(\Delta/T)(x+\delta p_{F}/\Delta)} \nonumber\\
&-&\int_{\delta p_{F}/\Delta}^{\infty}\frac{dx}{\sqrt{x^2-1}}
\sum_{n=1}^{\infty}(-1)^{n+1}{\rm e}^{-n(\Delta/T)(x-\delta p_{F}/\Delta)} \nonumber\\
&-& \int_{1}^{\delta p_{F}/\Delta} 
\frac{dx}{\sqrt{x^2-1}}\left(1-{\rm e}^{(\Delta/T)(x-\delta p_{F}/\Delta)}+...\right)
\label{eq:1.20}
\end{eqnarray}  
When $\delta p_{F}\leq\Delta$ the third integral is over the range
$[1,\infty)$, and there is no fourth integral. Since the second and third integrals
converge, we can set their upper limit equal to $\infty$.

For a not very large mismatch, $\delta p_{F}\sim \Delta$, we get 
\begin{equation}
\ln\left(\frac{\Delta_{0}}{\Delta}\right)=
2\sum_{n=1}^{\infty}(-1)^{n+1}
\cosh\left(\frac{n\delta p_{F}}{T}\right)K_{0}\left(\frac{n\Delta}{T}\right)
+\ln\left(\frac{\delta p_{F}+\sqrt{\delta p_{F}^2-\Delta^2}}{\Delta}\right)
\label{eq:1.21}
\end{equation}   
when $\delta p_{F}>\Delta$, 
while the last term on the right-hand side 
is absent when $\delta p_{F}\leq \Delta$. 
Here, we introduced the BCS gap at zero temperature,
$\Delta_{0}=\Delta_{0}(T=0)$, and we invoke 
the Bessel function
$K_{0}(z)=\int_{1}^{\infty}{\rm e}^{-zt}/\sqrt{t^2-1}\;dt$ \cite{GradshteynRyzhik}.

Using the asymptotic expansion of Bessel functions at large arguments
and solving Eq.~(\ref{eq:1.21}) we find 
\begin{eqnarray}
&& \Delta(T) = \left[\Delta_{0}(T)\;(2\delta p_{F}-\Delta_{0}(T))\right]^{1/2}
\qquad (\delta p_{F}>\Delta)
\nonumber\\
&& \Delta_{0}(T) = \Delta_{0}-\cosh\left(\frac{\delta p_{F}}{T}\right)\sqrt{2\pi T\Delta_{0}}
\left(1-\frac{T}{8\Delta_{0}}\right)
{\rm e}^{-(\Delta_{0}/T)}\qquad (\delta p_{F}\leq \Delta)
\label{eq:1.22}
\end{eqnarray}
where $\Delta(T)$ is the BP and $\Delta_{0}(T)$ is the BCS solutions
of the gap equation at low temperatures and finite Fermi momenta mismatch.
At $\delta p_{F}=0$ we recover the known dependence of the BCS gap
on the temperature \cite{Abrikosov}.

Next we calculate the condensation energy by integrating the gap equation,
Eq.~(\ref{eq:1.18}), over the gap parameter. 
We use the gap equation to eliminate $g$,
\begin{eqnarray}
&&\hspace{-0.5cm} \Omega_{BP}-\Omega_{N}= 2N(0)\int_{0}^{\Delta}\Delta^{\prime}d\Delta^{\prime}\left(
\ln\left(\frac{\Delta^{\prime}}{\Delta}\right)
-\ln\left(\frac{\Delta^{\prime}}{\Delta}\right)
+\ln\left(\frac{\delta p_{F}(\Delta^{\prime})
+\sqrt{\delta p_{F}(\Delta^{\prime})^{\,2}-\Delta^{\prime\,2}}}{\delta p_{F}(\Delta)
+\sqrt{\delta p_{F}(\Delta)^{\,2}-\Delta^{2}}}\right)
\right.\nonumber\\
&&\hspace{0.3cm} +\left. 2\sum_{n=1}^{\infty}(-1)^{n+1}
\left[\cosh\left(\frac{n\delta p_{F}(\Delta^{\prime})}{T}\right)
K_{0}\left(\frac{n\Delta^{\prime}}{T}\right)
-\cosh\left(\frac{n\delta p_{F}(\Delta)}{T}\right)
K_{0}\left(\frac{n\Delta}{T}\right)\right]\right)
\label{eq:1.22a}\end{eqnarray}  
Integrating by parts, we obtain
\begin{eqnarray}
\Omega_{BP}-\Omega_{N}&=& - N(0)\int_{0}^{\Delta}
\frac{\Delta^{\prime\,2}d\Delta^{\prime}}{\sqrt{\delta p_{F}(\Delta^{\prime})^2-\Delta^{\prime\,2}}}
\left[\frac{d\delta p_{F}(\Delta^{\prime})}{d\Delta^{\prime}}
-\frac{\Delta^{\prime}}{\delta p_{F}(\Delta^{\prime})+\sqrt{\delta p_{F}
(\Delta^{\prime})^2-\Delta^{\prime\,2}}}
\right]\nonumber\\
&+&N(0)\int_{0}^{\Delta}\Delta^{\prime\,2}d\Delta^{\prime}\frac{2}{T}
\sum_{n=1}^{\infty}(-1)^{n+1}n\left[\cosh\left(\frac{n\delta p_{F}(\Delta^{\prime})}{T}\right)
K_{1}\left(\frac{n\Delta^{\prime}}{T}\right)\right.\nonumber\\
&-&\left.\sinh\left(\frac{n\delta p_{F}(\Delta^{\prime})}{T}\right)
K_{0}\left(\frac{n\Delta^{\prime}}{T}\right)\frac{d\delta p_{F}(\Delta^{\prime})}{d\Delta^{\prime}}
\right]
\label{eq:0.1.22b}\end{eqnarray}
where we used the formula $K_{0}^{\prime}(z)=-K_{1}(z)$.
Using the parametrization $\delta p_{F}(\Delta)=\delta p_{F}(1+\alpha^2\frac{\Delta^2}{2\delta p_{F}^2})$,
we obtain in the leading order $O(\Delta^2/\delta p_{F}^2)$
\begin{eqnarray}
\Omega_{BP}-\Omega_{N} &=& -N(0)\frac{\Delta^{4}(2\alpha^2-1)}{8\delta p_{F}^2}
\nonumber\\
&+& N(0)2T^{\,2}\sum_{n=1}^{\infty}(-1)^{n+1}\left[\cosh\left(\frac{n\delta p_{F}}{T}\right)
\frac{1}{n^2}\int_{0}^{n\Delta/T}K_{1}(x)x^2dx\right.\nonumber\\
&-&\left.\frac{\alpha^2T}{\delta p_{F}}
\sinh\left(\frac{n\delta p_{F}}{T}\right)\frac{1}{n^3}
\int_{0}^{n\Delta/T}K_{0}(x)x^3dx\right]
\label{eq:1.22c}\end{eqnarray}
To evaluate the integrals, we write
$\int_{0}^{n\Delta/T}K_{1}(x)x^{2}dx=2-\int_{n\Delta/T}^{\infty}K_{1}(x)x^{2}dx$,
where in the remaining integral we use the asymptotic expansion of the function
$K_{1}$ (or equivalently we use the recursion relation for the Bessel functions
$\int x^{p+1}Z_{p}(x)dx=x^{p+1}Z_{p+1}$ \cite{GradshteynRyzhik});
and similar for the second integral. We obtain
\begin{eqnarray}
\int_{0}^{n\Delta/T}K_{1}(x)x^{2}dx &=&
2-\left(\frac{n\Delta}{T}\right)^{2}K_{2}\left(\frac{n\Delta}{T}\right)
\nonumber\\
\int_{0}^{n\Delta/T}K_{0}(x)x^{3}dx &=& 
4-\left[\left(\frac{n\Delta}{T}\right)^{3}K_{3}\left(\frac{n\Delta}{T}\right)
-2\left(\frac{n\Delta}{T}\right)^{2}K_{2}\left(\frac{n\Delta}{T}\right)\right]
\label{eq:1.22d}\end{eqnarray}   
where we used $K_{2}(z)=\frac{2}{z}K_{1}(z)+K_{0}(z)$,
$\int_{0}^{\infty}x^{\mu}K_{\nu}(ax)dx=2^{\mu-1}a^{-\mu-1}\Gamma\left(\frac{1+\mu+\nu}{2}\right)
\Gamma\left(\frac{1+\mu-\nu}{2}\right)$ \cite{GradshteynRyzhik}.
Since $\Delta/T\gg 1$ at low temperatures, the Bessel functions for $n=1$ give the dominant
contribution (we use the asymptotic expansion of the function $K_{\nu}(z)$).
For the constant term in Eq.~(\ref{eq:1.22d}), we need to calculate the sum in Eq.~(\ref{eq:1.22c}).
In the leading order $T/\Delta \ll 1$, we obtain
\begin{eqnarray}
&&\hspace{-1.cm} \Omega_{BP}-\Omega_{N} = -N(0)\frac{\Delta^{4}(2\alpha^2-1)}{8\delta p_{F}^2}
+N(0)\left[\phantom{\frac{1^1}{1^1}}\hspace{-0.4cm}2T^{2}\left(-{\rm Li}_{2}(-{\rm e}^{-(\delta p_{F}/T)})
-{\rm Li}_{2}(-{\rm e}^{(\delta p_{F}/T)})\right)
\right.\nonumber\\
&&\hspace{1.0cm}\left.-\cosh\left(\frac{\delta p_{F}}{T}\right)
\sqrt{2\pi\Delta^{3}T}\left(1+\frac{15}{8}\frac{T}{\Delta}\right){\rm e}^{-(\Delta/T)}
\right]\nonumber\\
&&\hspace{5.cm} -N(0)\frac{\alpha^2T}{\delta p_{F}}\left[\phantom{\frac{1^1}{1^1}}\hspace{-0.4cm}4T^{2}
\left({\rm Li}_{3}(-{\rm e}^{-(\delta p_{F}/T)})
-{\rm Li}_{3}(-{\rm e}^{(\delta p_{F}/T)})\right)
\right.\nonumber\\
&&\hspace{1.0cm}\left.-\sinh\left(\frac{\delta p_{F}}{T}\right)\left(\frac{\Delta}{T}\right)
\sqrt{2\pi\Delta^{3}T}\left(1+\frac{19}{8}\frac{T}{\Delta}\right){\rm e}^{-(\Delta/T)}
\right]
\label{eq:1.22e}\end{eqnarray} 
where the polylogarithmic functions are given by ${\rm Li}_{n}(z)=\sum_{k=1}^{\infty}z^{k}/k^{n}$,
$\frac{d}{dz}{\rm Li}_{n}(z)=\frac{1}{z}{\rm Li}_{n-1}(z)$, and
$\Delta$ is the BP solution of the gap equation given by 
Eq.~(\ref{eq:0.5}) with $\Delta_{0}$ defined in Eq.~(\ref{eq:1.22}).
When $\delta p_{F}=0$, terms in the first square bracket of Eq.~(\ref{eq:1.22e}) 
describe the temperature dependent part of the condensation energy in the BCS state 
\cite{Abrikosov}
\begin{eqnarray}
&&\hspace{-0.5cm} \Omega_{BCS}-\Omega_{N} = -N(0)\frac{\Delta_{0}^{2}}{2}
+N(0)\left[\frac{\pi^2T^2}{3}-\sqrt{2\pi\Delta_{0}^{3}T}\left(1+\frac{15}{8}
\frac{T}{\Delta_{0}}\right){\rm e}^{-(\Delta_0/T)}\right]
\label{eq:1.22g}\end{eqnarray} 
where the term $\sim T^{2}$ is the negative of the principal term in the expansion
of the free energy of the normal state in powers of $T$;
we used $\sum_{k=1}^{\infty}(-1)^{k+1}/k^2=\pi^2/12$. In Eq.~(\ref{eq:1.22e}), terms
dependent on the gap parameter contribute to the $\Omega_{BP}$, while both 
$\cosh$ and $\sinh$ terms give comparable contributions in the BP state
when $T\ll \Delta\sim \delta p_{F}$. This means that the BP heat capacity differes
from that of the BCS state and is modified by imposing
different physical conditions.     
      
{\it Large temperatures, $T\sim T_{c}$.}
To determine the behavior of the gap for the temperatures near the critical temperature
$T_{c}$, it is most convenient to start from the $4$-d relation Eq.~(\ref{eq:1.13}).
Near $T_{c}$ the size of the gap is small, and hence in Eq.~(\ref{eq:1.13}), we can carry
out an expansion in powers of $\Delta^2/T^2\ll 1$
\begin{eqnarray}
\frac{2}{gN(0)}&=& T\sum_{n=-\infty}^{\infty}\int_{-\omega}^{\omega}d\varepsilon
\left[\frac{1}{(\omega_{n}-i\delta p_{F})^2+\varepsilon^2}\right.\nonumber\\
 &-&\left. \frac{\Delta^2}{\left((\omega_{n}-i\delta p_{F})^2+\varepsilon^2\right)^2}
 +\frac{\Delta^{4}}{\left((\omega_{n}-i\delta p_{F})^2+\varepsilon^2\right)^3}+ ...\;
\right]
\label{eq:1.23}
\end{eqnarray}  
Interchanging the order of summation over the frequencies and integration over 
$\varepsilon$ in the (convergent) terms of the right-hand side, we obtain
\begin{eqnarray}
\frac{1}{gN(0)}&=& \int_{0}^{\omega}\frac{d\varepsilon}{\varepsilon}
\left(1-\frac{1}{{\rm e}^{(\varepsilon-\delta p_{F})/T}+1}
-\frac{1}{{\rm e}^{(\varepsilon+\delta p_{F})/T}+1}
\right) \label{eq:1.24} \\
&-& \frac{\Delta^2}{(\pi T)^2}\sum_{n=0}^{\infty}
\frac{1}{\left((2n+1)-i\frac{\delta p_{F}}{\pi T}\right)^3}
+\frac{3}{4}\frac{\Delta^4}{(\pi T)^4}\sum_{n=0}^{\infty}
\frac{1}{\left((2n+1)-i\frac{\delta p_{F}}{\pi T}\right)^5}
+...
\nonumber
\end{eqnarray}
To evaluate the integral, we split it into two pieces,
$\int_{0}^{\omega}dx=\int_{0}^{1}dx+\int_{1}^{\omega}dx$, where
$x=\varepsilon/T$, and we calculate each piece by doing necessary approximations
in the corresponding region.
Expressing the series appearing in Eq.~(\ref{eq:1.24}) in terms of the Riemann
zeta function, i.e. writing
$\sum_{n=0}^{\infty}\frac{1}{(2n+1)^{z}}=\frac{2^{z}-1}{2^{z}}\zeta(z)$, we find 
in the leading order
\begin{equation}
\ln\left(\frac{T}{T_{c}}\right)=
-2Ei(-1)\left[1-\cosh\left(\frac{\delta p_{F}}{T}\right)\right]
-\frac{7\zeta(3)}{8}\frac{\Delta^2}{(\pi T)^2}
\label{eq:1.25}
\end{equation}   
where the exponential-integral function is
$\int_{1}^{\infty}\frac{dx}{x}{\rm e}^{-\mu x}=-Ei(-\mu )$ \cite{GradshteynRyzhik}, 
$Ei(-1)\approx -0.219$ and $\zeta(3)\approx 1.202$,
and $T_{c}$ is the critical temperature for the BCS gap at zero mismatch.

Thus, in the leading order, we find that the size of the gap near $T_{c}$ is
\begin{eqnarray}
\Delta(T) = \Delta_{0}(T) &=& \pi T_{c}\sqrt{\frac{8}{7\zeta(3)}}\;
\sqrt{1-\frac{T}{T_{c}}
-2Ei(-1)\left[1-\cosh\left(\frac{\delta p_{F}}{T_{c}}\right)\right]}\nonumber\\
&\approx & 3.06T_{c}\sqrt{1-\frac{T}{T_{c}}
+0.44\left[1-\cosh\left(\frac{\delta p_{F}}{T_{c}}\right)\right]}
\qquad (\forall \delta p_{F})
\label{eq:1.26}
\end{eqnarray}  
for both the BP $(\delta p_{F}>\Delta)$ and the BCS $(\delta p_{F}\leq \Delta)$ 
pairing. At $\delta p_{F}=0$ we reproduce the known
temperature dependence of the BCS gap near $T_{c}$ \cite{Abrikosov}.
The behavior of the gap as a function of the Fermi momenta mismatch
for low  Eq.~(\ref{eq:1.22}) and high Eq.~(\ref{eq:1.26}) temperatures
is shown in Fig.(1) and Fig.(2), correspondingly. 
To parametrize $\Delta(T)$ we used the BCS parameters 
(at $\delta p_{F}=0$),
the gap at zero temperature $\Delta_{0}$
and the critical temperature $T_{c}$, which are related as
$T_{c}/\Delta_{0}={\rm e}^{\gamma_{E}}/\pi\approx 0.567$ where $\gamma_{E}\approx 0.577$
is the Euler's constant.

Integrating the gap equation Eq.~(\ref{eq:1.23}) over the gap parameter, we find
the condensation energy density at large temperatures
\begin{eqnarray}
\Omega_{BP}-\Omega_{N}&=& 2N(0)\int_{0}^{\Delta^{\prime}}\Delta^{\prime}d\Delta^{\prime}
\left(2(-Ei(-1))\left[\cosh\left(\frac{\delta p_{F}(\Delta^{\prime})}{T}\right)
-\cosh\left(\frac{\delta p_{F}(\Delta)}{T}\right)\right]\right.\nonumber\\
&+&\left.\frac{7\zeta(3)}{8}\frac{\Delta^{\prime\,2}-\Delta^{2}}{(\pi T)^2}
\right)
\label{eq:1.26a}\end{eqnarray} 
Integrating by parts, we obtain
\begin{eqnarray}
\Omega_{BP}-\Omega_{N}&=& -N(0)\frac{7\zeta(3)\Delta^4}{16(\pi T)^2}\nonumber\\
&-&N(0)\int_{0}^{\Delta}\Delta^{\prime\,2}d\Delta^{\prime}\frac{2(-Ei(-1))}{T}
\sinh\left(\frac{\delta p_{F}(\Delta^{\prime})}{T}\right)
\frac{d\delta p_{F}(\Delta^{\prime})}{d\Delta^{\prime}}
\label{eq:1.26b}\end{eqnarray} 
Parametrizing the Fermi momentum mismatch at large temperatures as 
$\delta p_{F}(\Delta)=\delta p_{F}(1+\beta^2\frac{\Delta^2}{2T^2})$,
we find from Eq.~(\ref{eq:1.26b})
\begin{eqnarray}
 \Omega_{BP}-\Omega_{N}&=& -N(0)\frac{\Delta^4}{T^2}\left[\frac{7\zeta(3)}{16\pi^2}
+\frac{\beta^2(-Ei(-1))}{2}\frac{\delta p_{F}}{T}
\sinh\left(\frac{\delta p_{F}}{T}\right)\right]\nonumber\\
&=&-\frac{2p_{F}^{2}T_{c}^{\,2}}{7\zeta(3)}\left[\left(1-\frac{T}{T_{c}}\right)^2
\left(1+\beta^2(-Ei(-1))\frac{8\pi^2}{7\zeta(3)}\frac{\delta p_{F}}{T_{c}}
\sinh\left(\frac{\delta p_{F}}{T_{c}}\right)\right)\right.\nonumber\\
&& +\left.\left(1-\frac{T}{T_{c}}\right)4(-Ei(-1))
\left(1-\cosh\left(\frac{\delta p_{F}}{T_{c}}\right)\right)\right] 
\label{eq:1.26c}\end{eqnarray}
where we used the gap equation solution, Eq.~(\ref{eq:1.26}), and kept the leading terms
$O(\delta p_{F}^2/T_{c}^2)$.
When $\delta p_{F}=0$ we reproduce the BCS free energy density, which is the leading term
in Eq.~(\ref{eq:1.26c}) since $\delta p_{F}\ll T_{c}$. 

Next,
we find the difference in the quark number densities in the breached pairing state 
at finite temperature.
For low temperatures, $\Delta/T\gg 1$, it is convenient to use the $3$-d formula,
Eq.~(\ref{eq:1.14}),
\begin{equation}
n_{u}-n_{s} = 2N(0)\;\Delta\int_{1}^{\omega/\Delta}\frac{xdx}{\sqrt{x^2-1}}\left[
\frac{1}{{\rm e}^{(\Delta/T)(x-\delta p_{F}/\Delta)}+1}
-\frac{1}{{\rm e}^{(\Delta/T)(x+\delta p_{F}/\Delta)}+1}
\right]
\label{eq:1.27}
\end{equation}    
where $x=\sqrt{\varepsilon^2+\Delta^2}/\Delta$, and $N(0)$ is the density
of states.
As with the gap equation, expanding the exponentials in the integrand, we find
in the leading order
\begin{equation}
n_{u}-n_{s} = 2N(0)\left[\sqrt{\delta p_{F}^2-\Delta^2}
+\sinh\left(\frac{\delta p_{F}}{T}\right)\sqrt{2\pi T\Delta_{0}}
\left(1+\frac{3T}{8\Delta_{0}}\right)
{\rm e}^{-(\Delta_{0}/T)}
\right] \label{eq:1.28}
\end{equation}  
when $\delta p_{F}>\Delta$.   
For $n_{u}-n_{s}$
we used integral representation of the appropriate Bessel function,
$K_{1}(z)=-\frac{d}{dz}K_{0}(z)=\int_{1}^{\infty}t{\rm e}^{-zt}/\sqrt{t^2-1}\;dt$ 
\cite{GradshteynRyzhik},
and its assymptotic behavior at large $z$ (see formula above).

For temperatures near critical, we use the $4$-d relation Eq.~(\ref{eq:1.15}).
As with the gap equation, expanding in powers of $\Delta^2/T^2$ and
performing summation over $n$ in the leading $\Delta$-independent term
and interchanging the order of summation
over the frequencies and integration over $\varepsilon$ in the convergent terms,
we obtain
\begin{eqnarray}
n_{u}-n_{s}&=& 2N(0) \left[\phantom{\frac{1^1}{\left(\frac{1}{1}\right)^2}}\hspace{-1.0cm}
\int_{0}^{\omega}d\varepsilon
\left(\frac{1}{{\rm e}^{(\varepsilon-\delta p_{F})/T}+1}
-\frac{1}{{\rm e}^{(\varepsilon+\delta p_{F})/T}+1}\right)\right.\nonumber\\
&-&\left. \frac{\Delta^2}{(\pi T)}\sum_{n=0}^{\infty}
\frac{1}{\left((2n+1)-i\frac{\delta p_{F}}{\pi T}\right)^2}+...\;
\right] 
\label{eq:1.29}\end{eqnarray}
Using the Riemann zeta function, we obtain in the leading order
\begin{eqnarray}
n_{u}-n_{s}&=& 2N(0)\left[T\left\{\frac{1}{2}\frac{\delta p_{F}}{T}
+\frac{2}{{\rm e}}\sinh\left(\frac{\delta p_{F}}{T}\right)\right\}
-\frac{3\zeta(2)}{4}\frac{\Delta^2}{(\pi T)}
\right]\nonumber\\
&\approx& 2N(0)\left[0.50\;\delta p_{F}+0.74\;T\sinh\left(\frac{\delta p_{F}}{T}\right)
-0.39\;\frac{\Delta^2}{T}\right]
\label{eq:1.30}
\end{eqnarray} 
where $\zeta(2)=\pi^2/6$. 
Sum of the quark number densities at nonzero temperature coincide with that
at zero termperature, Eq.~(\ref{eq:0.11a}),  
\begin{eqnarray}
n_{u}+n_{s}=2N(0)\,\frac{p_{F}}{3}
\label{eq:0.11ab}\end{eqnarray} 
since the temperature dependent term in Eq.~(\ref{eq:1.14})
is an odd function of $\varepsilon$. Particle number density of a free gas 
gets the temperature correction, Eq.~(\ref{eq:1.22g}), for example 
\begin{eqnarray}
n_{u}=\frac{p_{F}^{u}}{3\pi^2}\left(p_{F}^{u\,2}+\pi^2 T^2\right)
\label{eq:0.11abc}\end{eqnarray}   

In the following section,  
we use the BP solutions of the gap equation at zero, Eq.~(\ref{eq:0.3b}),
and nonzero temperatures,
Eqs.~(\ref{eq:1.22},\ref{eq:1.26}), as well as
the quark number densities at $T=0$, Eq.~(\ref{eq:0.11a}), and at $T\neq 0$, 
Eqs.~(\ref{eq:1.28},\ref{eq:1.30},\ref{eq:0.11ab},\ref{eq:0.11abc}),
to find the breached pairing states of the quark matter which are electrically neutral.

\section{Breached pairing in electrically neutral quark matter}

In this section we solve the gap equation(s) imposing condition
of the electric neutrality. In addition, when we require the neutrality 
with respect to the color charges the energy minimum shifts slightly \cite{ShovkovyHuang}.
We therefore put
the chemical potentials associated
with color charges of the $U(1)_{3}\times U(1)_{8}$ subgroup to zero, 
$\mu_{3}=\mu_{8}=0$. This approximation does not change our results qualitatively. 

In the color and flavor antisymmetric channel, the three flavor gap parameter is parametrized
by three gaps, $\Delta_{ud},\,\Delta_{us},\,\Delta_{ds}$, (Appendix A).
One minimizes the thermodynamic potential $\Omega$,
\begin{eqnarray}
\frac{\partial\Omega}{\partial \Delta_{i}}=0,\qquad
\frac{\partial\Omega}{\partial \mu_{e}}=0
\label{eq:2.0}\end{eqnarray}  
with respect to the gap parameters $\Delta_{i}$ and the electric chemical potential
$\mu_{e}$ associated with the electric $U(1)$ charge. This is equivalent
to looking for the minimum of $\Omega$ in the $(\Delta_{i},\,\mu_{e})$ plane
only along the electric neutrality line \cite{ShovkovyHuang}. 
We however do not solve directly minimization problem for $\Omega$.
Numerical minimization was pursued for two flavors in \cite{ShovkovyHuang}
and for three flavors in \cite{AlfordKouvarisRajagopal}.
Here, we solve analytically the gap equations and the neutrality conditions
and find the parameter space where they intersect.

Expressed through the quark number densities,
electric neutrality requires 
$\frac{2}{3}n_{u}-\frac{1}{3}n_{d}-\frac{1}{3}n_{s}-n_{e}=0$,
or written explicitly for each pair,  
\begin{eqnarray}
&&\hspace{-2cm} \frac{1}{2}\left(\frac{2}{3}n_{u}^{(1)}-\frac{1}{3}n_{d}^{(2)}
+\frac{2}{3}n_{u}^{(1)}-\frac{1}{3}n_{s}^{(3)}
-\frac{1}{3}n_{d}^{(2)}-\frac{1}{3}n_{s}^{(3)}\right) \nonumber\\
&&\hspace{-2cm} +\left(\frac{2}{3}n_{u}^{(2)}-\frac{1}{3}n_{d}^{(1)}
+\frac{2}{3}n_{u}^{(3)}-\frac{1}{3}n_{s}^{(1)}
-\frac{1}{3}n_{d}^{(3)}-\frac{1}{3}n_{s}^{(2)}\right)
-n_{e}=0
\label{eq:2.1}\end{eqnarray} 
where the upper index $1,2,3$ denotes color red, green, blue, correspondingly,
the first brackets contains quarks from the $3\times 3$ block
and the second bracket from the $2\times 2$ blocks (Appendix A).
The quark densities are defined by the quark Fermi momenta, given by
$p_{F}^{u}=\mu-(2/3)\mu_{e},\;p_{F}^{d}=\mu+(1/3)\mu_{e},\;
p_{F}^{s}=\mu+(1/3)\mu_{e}-m_{s}^{2}/2\mu$
where $\mu$ is the baryon chemical potential and $\mu_{e}$ 
is the electric chemical potential.
For convenience we introduce the average and the mismatch in Fermi momenta  
for each pair, for example, for  
$p_{ud}=(p_{F}^{u}+p_{F}^{d})/2$
and $\delta p_{ud}=(p_{F}^{d}-p_{F}^{u})/2$ 
where $\delta p$ is chosen to be positive.
Explicitly, we have 
\begin{eqnarray}
p_{ud} &=& \mu-\frac{\mu_{e}}{6} \hspace{2.09cm} 
\delta p_{ud}=\frac{\mu_{e}}{2} \nonumber\\
p_{us} &=& \mu-\frac{\mu_{e}}{6}-\frac{m_{s}^{2}}{4\mu} \hspace{1cm}
\delta p_{us}=\frac{m_{s}^{2}}{4\mu}-\frac{\mu_{e}}{2} \nonumber\\
p_{ds} &=& \mu+\frac{\mu_{e}}{3}-\frac{m_{s}^{2}}{4\mu} \hspace{0.98cm}
\delta p_{ds}=\frac{m_{s}^{2}}{4\mu}
\label{eq:2.2}\end{eqnarray}
where the ordering of the Fermi momenta in the non-interacting electrically neutral 
quark medium is \cite{AlfordRajagopal}
$p_{F}^{s}<p_{F}^{u}<p_{F}^{d}$.
In what follows we specify the quark number densities and 
solve the gap equations under condition of the electric neutrality
for two and three flavors.

\subsection{Two and three flavor stable quark matter at zero temperature}

{\it Three flavor mixed BCS-BP state.}
At asymptotically high densities three flavor quark matter is in the CFL phase.
Going to lower densities, the effective strange quark mass increases
making impossible to maintain the BCS pairing for all nine quarks (the CFL phase).
The CFL phase becomes unstable when first the down-strange quarks break the BCS pairing
at $m_{s}^2/4\mu\geq \Delta_{0}$, Eq.~(\ref{eq:0.3b},\ref{eq:2.2}),
where $\Delta_{0}$ is the CFL gap.

We find that,
due to the small Fermi momentum of the strange quark,
$p_{F}^{s}\ll p_{F}^{u}< p_{F}^{d}$, 
up blue-strange red and down blue-strange green quarks participate 
in the breached pairing
while up green-down red quarks pair according to the BCS mechanism.
The remaining up red, down green and strange blue quarks 
form the BCS condensate. Out of total nine quasiparticle excitations, there are
five BCS and four BP modes, leading to seven gapped and two gapless excitations (Appendix A).
The system of gap equations for the BCS $\Delta_{ud}$
and the BP $\Delta_{us}\sim\Delta_{ds}$ gap parameters decouples in the $\bar{3}$ channel
and reduces to the two two-flavor gap equations (Appendix A). We therefore use 
the quark number densities and the stability criterion obtained for the two-flavor case
in the previous sections.  

When quarks participate in the BCS pairing, their Fermi momenta and corresponding 
densities are modified 
by the strong interactions in a way to equalize the densities
of two species which form the Cooper pairs \cite{RajagopalWilczek}.
In our case, $n_{u}^{(1)}=n_{d}^{(2)}=n_{s}^{(3)}$,
meaning that up red, down green and strange blue quarks form an electrically neutral
quark matter, $\frac{2}{3}n_{u}^{(1)}-\frac{1}{3}n_{d}^{(2)}
+\frac{2}{3}n_{u}^{(1)}-\frac{1}{3}n_{s}^{(3)}-\frac{1}{3}n_{d}^{(2)}-\frac{1}{3}n_{s}^{(3)}=0$.
Therefore the electric neutrality condition, Eq.~(\ref{eq:2.1}), including quarks of other colors
is given by 
\begin{equation}
\left(\frac{2}{3}n_{u}-\frac{1}{3}n_{d}\right)_{BCS}
+\left(\frac{2}{3}n_{u}-\frac{1}{3}n_{s}
-\frac{1}{3}n_{d}-\frac{1}{3}n_{s}\right)_{BP}
-n_{e}=0
\label{eq:3.1}\end{equation} 
where $n_{e}=\mu_{e}^{3}/(3\pi^2)$.
Using Eq.~(\ref{eq:0.11a}) for the quark number densities, 
we obtain
\begin{eqnarray}
&& \frac{1}{3}(\mu-\frac{\mu_{e}}{6})^{3}
+(\mu-\frac{\mu_{e}}{6}-\frac{m_{s}^2}{4\mu})^{2}\left[
\frac{1}{3}(\mu-\frac{\mu_{e}}{6}-\frac{m_{s}^2}{4\mu})
+3\sqrt{(\frac{m_{s}^2}{4\mu}-\frac{\mu_{e}}{2})^{2}-\Delta^{2}}
\,\right]\nonumber\\
&& \hspace{2.2cm} -\frac{2}{3}(\mu+\frac{\mu_{e}}{3}-\frac{m_{s}^2}{4\mu})^{3}-\mu_{e}^{3}=0
\label{eq:3.2}\end{eqnarray} 
where $\Delta\equiv \Delta_{us}$.
Since down and strange quarks carry the same electric charge, the breached region (with no pairing)
from the $\langle ds\rangle$ condensate does not contribute to the electric neutrality.
The breached region from the $\langle us\rangle$ condensate contribute to the excess
of the positive electric charge, that already exists in the non-interacting quark matter
where strange quarks are less abandon than up and down quarks at $m_{s}\neq 0$.
Therefore a finite density of electrons in needed, $\mu_{e}\neq 0$, to satisfy
the electric neutrality. 

Since Eq.~(\ref{eq:3.2}) includes the gap parameter $\Delta_{us}$, we solve
the neutrality condition with respect to $\delta p_{us}=(m_{s}^2/4\mu-\mu_{e}/2)$ 
before ($\Delta_{us}=0$) and after ($\Delta_{us}\neq 0$) pairing.
Using the parametrization $\delta p_{us}(\Delta_{us})=\delta p_{us}
(1+\alpha^2\frac{\Delta_{us}^2}{2\delta p_{us}^2})$, we obtain in the leading order
\begin{equation}
\alpha^2=\frac{3(3-4R+x)^{2}}{45+271R^2+24x+231x^2-2R(45+239x)}
\label{eq:3.3}\end{equation} 
where $x=\frac{\delta p_{us}}{\mu}$ is the numerical solution of the neutrality
condition Eq.~(\ref{eq:3.2}) before pairing at some fixed strange quark mass 
and $R=\frac{m_{s}^2}{4\mu^2}$. 
According to the stability criteria, Eqs.~(\ref{eq:0.5},\ref{0.10}), 
the BP solution is stable when $\alpha^2 > 1/2$. 
Fixing the baryon chemical potential $\mu=400\,MeV$ and 
increasing the strange quark mass
$m_{s}=150-327\,MeV$, we obtain $|p_{F}^{u}-p_{F}^{s}|=14-56\,MeV$ which grows 
and the coefficient in the stability criteria for the BP $\Delta_{us}$ solution 
$\alpha^2=0.59-0.50$ which drops. Satisfying the condition of breaking
the CFL state, $\Delta_{0}<m_{s}^2/4\mu$ with
$\Delta_{0}=20\,MeV$, we have $m_{s}>179 \,MeV$.   
This means that in this range of parameters, $179< m_{s}< 327\,MeV$
and under the neutrality condition Eq.~(\ref{eq:3.2})    
the mixed BCS-BP phase for three flavors is stable, while
for larger masses, $m_{s}> 327\,MeV$ it becomes unstable.

The similar phase, containing seven gapped and two gappless excitations, 
was obtained numerically by Alford, Kouvaris, Rajagopal 
\cite{AlfordKouvarisRajagopal}, and was called by the authors the gapless CFL.

When the ordering of the Fermi momenta in the strongly interacting neutral quark matter
follows the pattern $p_{F}^{s}\ll p_{F}^{u}< p_{F}^{d}$, there is a hirarchi of scales
between the gap parameters $\Delta_{ds}<\Delta_{us}<\Delta_{ud}$. We might 
find phases containing six (when down and strange quarks do not pair) and eight 
BP modes. The phase with the maximum number (eight) of the BP modes requires
the comparable Fermi momenta mismatches between different pairs and is the BP analog
of the CFL phase. It seems, however, that these phases are not stable
under condition of the electric neutality at zero temperature.

{\it Two flavor BP state.}
At large strange quark mass, only up and down quarks (up red with down green,
and up green with down red) participate in the breached pairing,
while strange quarks of all three colors and blue up and down quarks 
form the free gas. The electric neutrality condition, Eq.~(\ref{eq:2.1}),  
is written
\begin{eqnarray}
2\left(\frac{2}{3}n_{u}-\frac{1}{3}n_{d}\right)_{BP}
+\frac{2}{3}n_{u}-\frac{1}{3}n_{d}-\frac{1}{3}3n_{s}-n_{e}=0
\label{eq:3.8}\end{eqnarray} 
Using the quark number densities, we obtain
\begin{eqnarray}
&&\hspace{-1cm} 2(\mu-\frac{\mu_{e}}{6})^{2}\left[\frac{1}{3}(\mu-\frac{\mu_{e}}{6})
-3\sqrt{(\frac{\mu_{e}}{2})^{2}-\Delta_{ud}^{2}}\,\right]
 +\frac{2}{3}(\mu-\frac{2\mu_{e}}{3})^{3}-\frac{1}{3}(\mu+\frac{\mu_{e}}{3})^{3}
\nonumber\\
&&\hspace{3.3cm} -(\mu+\frac{\mu_{e}}{3}-\frac{m_{s}^2}{2\mu})^{3}-\mu_{e}^{3}=0
\label{eq:3.9}\end{eqnarray} 
where $\Delta\equiv \Delta_{ud}$. 
We solve
the neutrality condition, Eq.~(\ref{eq:3.9}), 
with respect to $\delta p_{ud}=\mu_{e}/2$ 
before ($\Delta_{ud}=0$) and after ($\Delta_{ud}\neq 0$) pairing.
Using the parametrization $\delta p_{ud}(\Delta_{ud})=\delta p_{ud}
(1+\alpha^2\frac{\Delta_{ud}^2}{2\delta p_{ud}^2})$, we obtain in the leading order
\begin{eqnarray}
\alpha^2=\frac{9-6x+x^2}{18-18x+48x^2+12R^2-4R(3+2x)}
\label{eq:3.10}\end{eqnarray} 
and for the infinitely large $m_{s}$ when the free gas of the strange quarks 
does not participate in the neutrality balance,
\begin{eqnarray}
\alpha^2=\frac{27-18x+3x^2}{45-66x+140x^2}
\label{eq:3.11}\end{eqnarray} 
Increasing the strange quark mass $m_{s}=200-400\,MeV$,
we obtain the electric chemical potential $\mu_{e}=23-70\,MeV$
which grows and the coefficient for the stability criterion is
$\alpha^2=0.53-0.59$ which also grows. Satisfying 
$\Delta_{0}<m_{s}^2/4\mu$ with $\Delta_{0}=50\,MeV$,
we have $m_{s}>283\,MeV$. For two flavors we need higher coupling ($\Delta_{0}$)
than for three flavor case in order to satisfy the electric neutrality condition,
i.e. the electric neutrality point $\mu_{e}(\Delta=0)$ should be to the left
from the gap equation point $\mu_{e}(\Delta=0)=\Delta_{0}$ (see Fig.(1)).
For $m_{s}\rightarrow \infty$, we obtain $\mu_{e}=86\,MeV$
and $\alpha^2=0.64$. Therefore the BP two-flavor phase is stable
for $m_{s}>283\,MeV$, and there is no restriction on $m_{s}$
from above. 

At $m_{s}\rightarrow\infty$, this phase was obtained numerically 
by Shovkovy, Huang at $T=0$ \cite{ShovkovyHuang} and $T\neq 0$ 
\cite{HuangShovkovy}, \cite{LiaoZhuang}
and was called by the authors the gappless 2SC.

\subsection{Stability of the quark matter at nonzero temperature}

Here we find the solution of the gap equation under the condition of the electric neutrality
at nonzero temperature graphically. We use the gap equation solutions, 
Eqs.~(\ref{eq:1.22},\ref{eq:1.26}), and the quark number densities,
Eqs.~(\ref{eq:1.28},\ref{eq:1.30},\ref{eq:0.11ab},\ref{eq:0.11abc}), at nonzero temperature.

\vspace{1cm}

\begin{figure}[htbp]\unitlength1cm
\begin{picture}(3,7)
\put(0,1){\epsfig{file=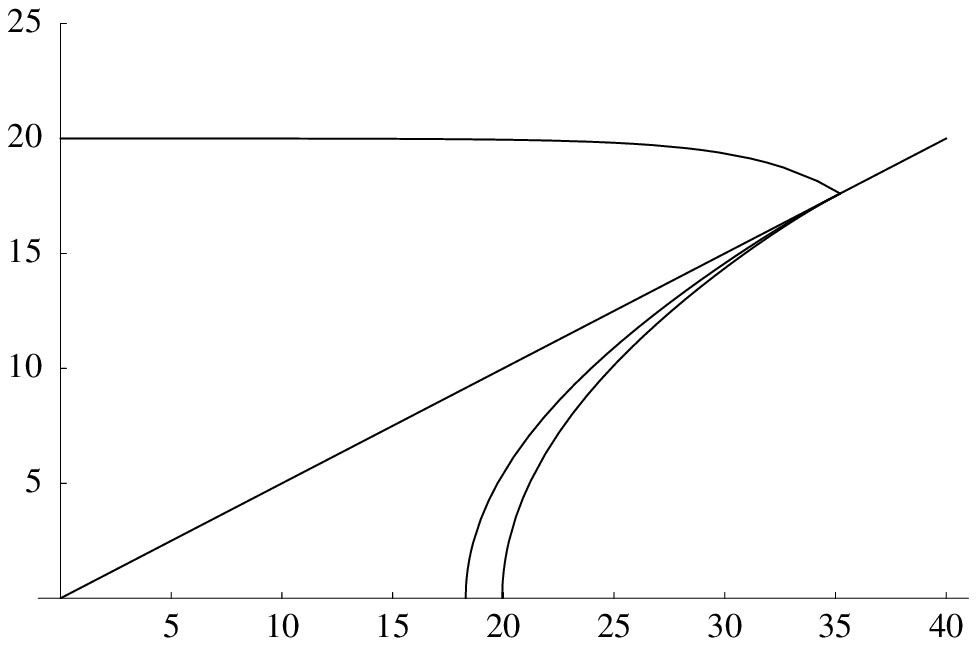,width=7cm,height=6cm}}
\put(8,1){\epsfig{file=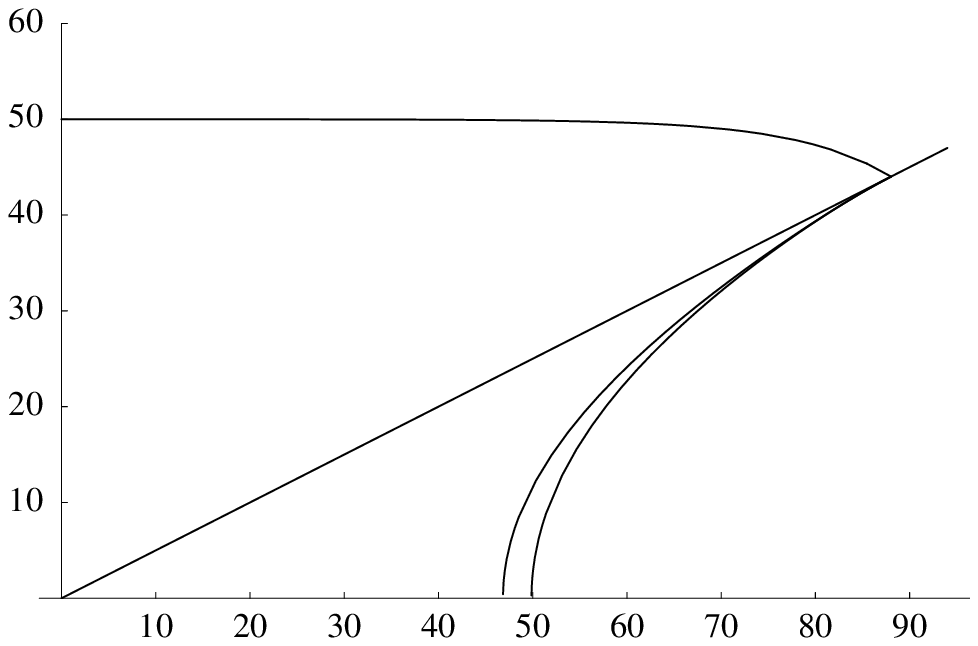,width=7cm,height=6cm}}
\put(3.2,0.5){ $\left[m_s^2/2\mu-\mu_{e}\right]\,$(MeV)}
\put(13,0.5){ $\mu_{e}\,$(MeV)}
\put(0,7.5){$\Delta\,$(MeV)}
\put(8,7.5){$\Delta\,$(MeV)}
\put(3,6.2){BCS}
\put(11,6.2){BCS}
\put(4.,2){BP}
\put(12.2,2){BP}
\put(2.8,2){N}
\put(11,2){N}
\end{picture}
\caption{Gap as a function of the Fermi momenta mismatch, $2\delta p_{F}$,
at small temperatures.
The BCS and BP are the two branches of the gap equation solution, and N denotes 
the electric neutrality line. Left: Three flavor mixed BCS-BP state,
$T=2\,MeV$, $\Delta_{0}=20\,MeV$, $m_{s}=260\,MeV$.
Right: Two flavor BP state,
$T=5\,MeV$, $\Delta_{0}=50\,MeV$, $m_{s}=300\,MeV$.
In both cases $\mu=400\,MeV$.}
\end{figure}

{\it Three flavor mixed BCS-BP state at $T\neq 0$.}
As at zero temperature, we analyze the BP gaps $\Delta_{us}\sim\Delta_{ds}$
without reference to the BCS gaps formed by the other quarks (Appendix A).
We specify the electric neutrality for the BCS-BP state,  
Eq.~(\ref{eq:3.1}), at small temperatures
\begin{eqnarray}
&&\hspace{-1cm} \frac{1}{3}\left(\mu-\frac{\mu_{e}}{6}\right)^3
+\left(\mu-\frac{\mu_{e}}{6}-\frac{m_{s}^2}{4\mu}\right)^2\left[
\frac{1}{3}\left(\mu-\frac{\mu_{e}}{6}-\frac{m_{s}^2}{4\mu}\right)
+3\sqrt{\left(\frac{m_{s}^2}{4\mu}-\frac{\mu_{e}}{2}\right)^2-\Delta^2}\right.\label{eq:4.1}\\
&&\hspace{-1cm}+\left.3\sinh\left(\frac{1}{T}
\left(\frac{m_{s}^2}{4\mu}-\frac{\mu_{e}}{2}\right)\right)\sqrt{2\pi T\Delta_{0}}\;
{\rm e}^{-\left(\Delta_{0}/T\right)}\right]
-\frac{2}{3}\left(\mu+\frac{\mu_{e}}{3}-\frac{m_{s}^2}{4\mu}\right)^3 
-\mu_{e}\left(\mu_{e}^2+\pi^2T^2\right)=0\nonumber
\end{eqnarray}   
and large temperatures, 
\begin{eqnarray}
&&\hspace{-2cm}\frac{1}{3} \left(\mu-\frac{\mu_{e}}{6}\right)^3
+\left(\mu-\frac{\mu_{e}}{6}-\frac{m_{s}^2}{4\mu}\right)^2\left[
\frac{1}{3}\left(\mu-\frac{\mu_{e}}{6}-\frac{m_{s}^2}{4\mu}\right)\right.\nonumber\\
&&\hspace{-2.cm}+\left.3T\left\{\frac{1}{2T}\left(\frac{m_{s}^2}{4\mu}-\frac{\mu_{e}}{2}\right)
+\frac{2}{{\rm e}}\sinh\left(\frac{1}{T}
\left(\frac{m_{s}^2}{4\mu}-\frac{\mu_{e}}{2}\right)\right)
\right\}-\frac{3\pi}{8}\frac{\Delta^2}{T}\right]\nonumber\\
&&\hspace{-2.cm}-\frac{2}{3}\left(\mu+\frac{\mu_{e}}{3}-\frac{m_{s}^2}{4\mu}\right)^3 
-\mu_{e}\left(\mu_{e}+\pi^2T^2\right)=0
\label{eq:4.2}\end{eqnarray}
We solve the neutrality condition with respect to $\Delta$ as a function of
the Fermi momenta mismatch.   
In Fig.(1), the neutrality line first intersects the gap equation solution
and then the border between the BP and the BCS states, given by $\Delta=\delta p_{F}$.
Quark matter is positively charged to the left from the neutrality line,
and it is negatively charged to the right. In order to get an intersection
between the gap equation solution and the neutrality line, the Fermi momenta mismatch
in the neutrality condition for a free gas should satisfy
$\delta p_{F}(\Delta=0)<\Delta_{0}$ where $\Delta_{0}$ is the BCS gap.
We therefore have stronger coupling for the two flavors ($\Delta_{0}=50\,MeV$)
than for the three flavors ($\Delta_{0}=20\,MeV$). This agrees with the renormalization group scaling
of the coupling constant as we go from high to lower densities.
We obtain the electrically neutral solution in the range of strange quark masses
which agrees with our stability analyses at $T=0$.   
      
\vspace{1cm}

\begin{figure}[htbp]\unitlength1cm
\begin{picture}(3,7)
\put(0,1){\epsfig{file=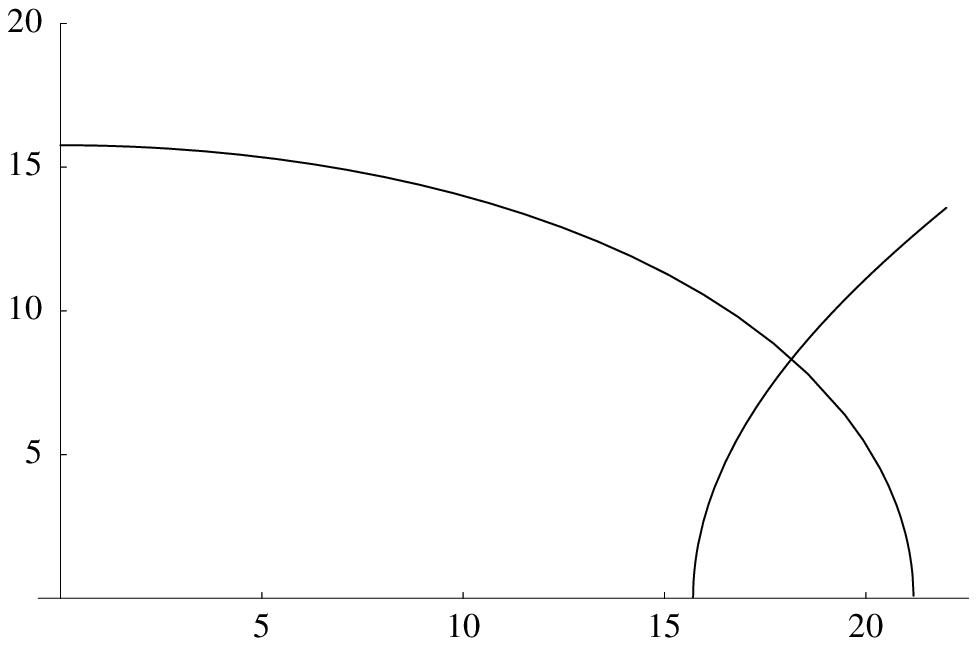,width=7cm,height=6cm}}
\put(8,1){\epsfig{file=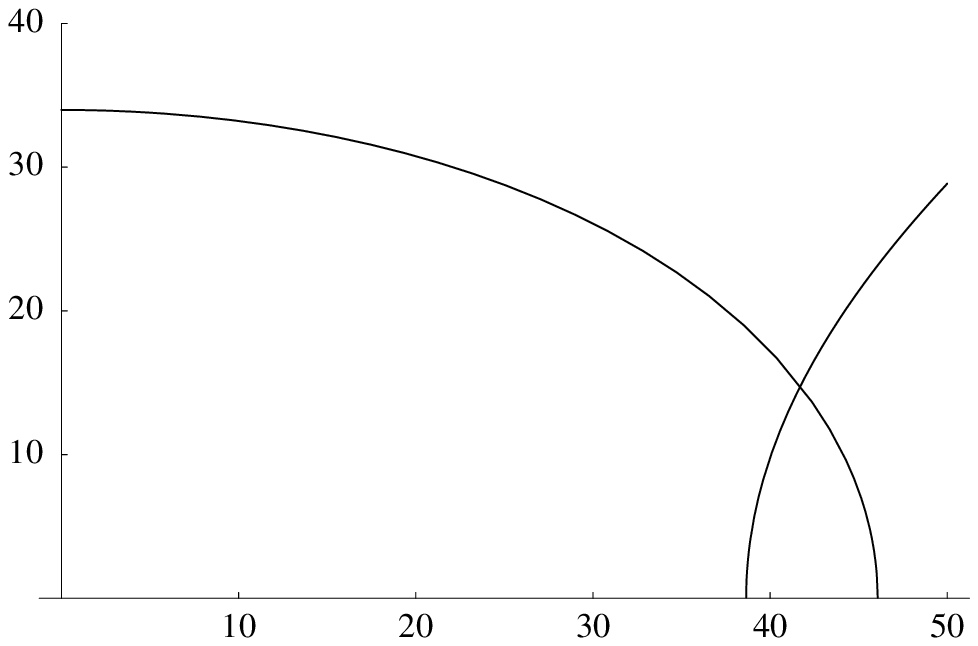,width=7cm,height=6cm}}
\put(3,0.5){ $\left[m_s^2/2\mu-\mu_{e}\right]\,$(MeV)}
\put(13,0.5){ $\mu_{e}\,$(MeV)}
\put(0.,7.5){$\Delta\,$(MeV)}
\put(8,7.5){$\Delta\,$(MeV)}
\put(2.4,6){BCS\,and\,BP}
\put(11,6){BCS\,and\,BP}
\put(4.3,2){N}
\put(12.7,2){N}
\end{picture}
\caption{Gap as a function of the Fermi momenta mismatch, $2\delta p_{F}$,
at large temperatures.
Left: Three flavor mixed BCS-BP state,
$T=9\,MeV$, $T_{c}=11\,MeV$, $m_{s}=260\,MeV$.
Right: Two flavor BP state,
$T=24\,MeV$, $T_{c}=28\,MeV$, $m_{s}=300\,MeV$.
In both cases $\mu=400\,MeV$.}
\end{figure}

{\it Two flavor BP state at $T\neq 0$.}
As at zero temperature, we consider the BP gap $\Delta_{ud}$.
The electric neutrality for the two flavor BP state,  
Eq.~(\ref{eq:3.8}), is written at small temperatures
\begin{eqnarray}
&&\hspace{-1cm}2\left(\mu-\frac{\mu_{e}}{6}\right)^2\left[
\frac{1}{3}\left(\mu-\frac{\mu_{e}}{6}\right)
-3\sqrt{\left(\frac{\mu_{e}}{2}\right)^2-\Delta^2}
-3\sinh\left(\frac{\mu_{e}}{2T}\right)\sqrt{2\pi T\Delta_{0}}\;
{\rm e}^{-\left(\Delta_{0}/T\right)}\right]\nonumber\\
&&\hspace{-1cm}+\frac{2}{3}(\mu-\frac{2\mu_{e}}{3})\left((\mu-\frac{2\mu_{e}}{3})^2+\pi^2T^2\right)
-\frac{1}{3}(\mu+\frac{\mu_{e}}{3})\left((\mu+\frac{\mu_{e}}{3})^2+\pi^2T^2\right)\nonumber\\
&&\hspace{-1cm}-(\mu+\frac{\mu_{e}}{3}-\frac{m_{s}^2}{2\mu})
\left((\mu+\frac{\mu_{e}}{3}-\frac{m_{s}^2}{2\mu})^2+\pi^2T^2\right)
-\mu_{e}\left(\mu_{e}^2+\pi^2T^2\right)=0 
\label{eq:4.3}\end{eqnarray}    
and at large temperatures, 
\begin{eqnarray}
&&\hspace{-1cm}2\left(\mu-\frac{\mu_{e}}{6}\right)^2\left[
\frac{1}{3}\left(\mu-\frac{\mu_{e}}{6}\right)
-3T\left\{\frac{\mu_{e}}{4T}+\frac{2}{{\rm e}}\sinh\left(\frac{\mu_{e}}{2T}\right)\right\}
+\frac{3\pi}{8}\frac{\Delta^2}{T}\right]\nonumber\\
&&\hspace{-1cm}+\frac{2}{3}(\mu-\frac{2\mu_{e}}{3})\left((\mu-\frac{2\mu_{e}}{3})^2+\pi^2T^2\right)
-\frac{1}{3}(\mu+\frac{\mu_{e}}{3})\left((\mu+\frac{\mu_{e}}{3})^2+\pi^2T^2\right)\nonumber\\
&&\hspace{-1cm}-(\mu+\frac{\mu_{e}}{3}-\frac{m_{s}^2}{2\mu})
\left((\mu+\frac{\mu_{e}}{3}-\frac{m_{s}^2}{2\mu})^2+\pi^2T^2\right)
-\mu_{e}\left(\mu_{e}^2+\pi^2T^2\right)=0
\label{eq:4.4}\end{eqnarray}
In Fig.(2) we used the same set of parameters as in Fig.(1) (including $T_{c}=0.567\Delta_{0}$), 
but only increased the temperature. 
At temperatures near the critical one, the BCS and the BP solutions 
form one curve. We find that it is much simpler to satisfy
the neutrality condition at high temperatures, that leads to a larger parameter
space where the neutral BP phase is possible.    
High temperatures also open an opportunity for neutral phases contaning
more than four BP modes.

\section{Conclusions}    

We analyzed the breached pairing superconductivity at zero and finite temperatures.
As in the previous work \cite{GubankovaLiuWilczek}, we found that the additional constraints
and physical conditions are crucial for the stability of the BP phase.
We found analytical expressions for the stability criteria, showing the parameter space
where the BP phase is stable under a general constraint.

Imposing the condition of the electric neutrality, we found the two-flavor
BP and the three flavor mixed BCS-BP phases, which are stable over a wide range of
parameters. Both phases contain four BP modes in their quasiparticle spectrum.
Fixing the chemical potential at $\mu=400\,MeV$ and increasing
the effective strange quark mass, we found that the CFL breaks at $m_{s}\sim 179\,MeV$,
followed by BP phases including the mixed BCS-BP phase for $179<m_{s}<327\,MeV$ and 
the two flavor BP phase for $283<m_{s}$. Phases containing more than four BP modes
which preclude the two-flavor BP phase might be also possible. 
In the region of $m_{s}$ where different BP phases overlap, a phase which is more 
energetically favorable wins.

At nonzero temperature we found solutions of the gap equation,
which are consistent with the numerically found solutions in \cite{Sarma}.
At low temperatures, there are two distinct branches, the BCS and the BP solutions,
while there is only one curve for both solutions at high temperatures. 
It is much simpler to satisfy the neutrality condition at high temperatures, 
leading to a larger parameter
space where the neutral BP phases are possible.    
High temperatures also open an opportunity for neutral phases contaning
more than four BP modes.

\section*{Acknowledgements}

The author would like to thank Frank Wilczek for his insight and useful
suggestions over the course of this work and for reading the manuscript;
and Michael Forbes, Chris Kouvaris and Krishna Rajagopal for many helpful
discussions.
This work is supported in part by funds provided by the U.S.\ Department of
Energy under cooperative research agreement DF-FC02-94ER40818.

\section*{Appendix. Gap equations for two and three flavors}

In this appendix, we obtain the gap equations for diquark condensates
containing two and three quark flavors. We use 
a toy model where 
the full interaction between quarks is replaced by a four-fermion interaction with
the quantum numbers of a single-gluon exchange,
${\it L}_{int}=G\int d^{4}x({\bar \psi}(x)_{i}^{\alpha}\lambda^{A}_{\alpha\beta}
\gamma^{\mu}\psi(x)_{i}^{\beta})
({\bar \psi}(y)_{j}^{\gamma}\lambda^{A}_{\gamma\delta}\gamma_{\mu}\psi(y)_{j}^{\delta})$,
where $\alpha, \beta$, etc. are color indices and $i,j$ are the flavor indices,
$\lambda^{A}$ are the color $SU(3)$ generators satisfying
$\lambda^{A}_{\alpha\beta}\lambda^{A}_{\gamma\delta}
=\frac{2}{3}(3\delta_{\alpha\delta}\delta_{\gamma\beta}
-\delta_{\alpha\beta}\delta_{\gamma\delta})$.
We allow condensation in the channel 
$\Delta_{ij}^{\alpha\beta}=G\langle\psi_{i}^{\alpha}C\gamma_{5}\psi_{j}^{\beta}
\rangle$ with the simplest color-flavor structure,
suggested in \cite{AlfordBergesRajagopal}, 
that interpolates
between the color-flavor locking phase at $M_{s}=0$ and the 2SC at large $M_{s}$,
\begin{equation}
\Delta^{\alpha\gamma}_{ij}=
\left(\begin{array}{ccccccccc}
b+e&b&c&&&&&&\\
b&b+e&c&&&&&&\\
c&c&d&&&&&&\\
&&&&e&&&&\\
&&&e&&&&&\\
&&&&&&f&&\\
&&&&&f&&&\\
&&&&&&&&f\\
&&&&&&&f&
\end{array}\right)
\,,\label{eq:a.0}\end{equation}
where the basis vectors are 
$(\alpha, i)=(1, 1), (2, 2), (3, 3), (1, 2), (2, 1), (1, 3), (3, 1), (2, 3), (3, 2)$
with colors $(1, 2, 3)$ correspond to red, green, blue and flavors $(1, 2, 3)$
correspond to up, down, strange \cite{AlfordBergesRajagopal}.
Detailed properties of the condensate ansatz are discussed in \cite{AlfordBergesRajagopal},
in particular this condensate locks color and flavor.   
In the mean field, the four-fermion interaction ${\it L}_{int}$
leads to an effective action
\begin{equation}
(\psi^{\dagger} \psi)
\left(\begin{array}{cc}
p_{0}I-E& Q\\
Q& p_{0}I+E
\end{array}\right)
\left(\begin{array}{c}
\psi\\
\psi^{\dagger}
\end{array}\right)
\label{eq:a.1}\end{equation}
where $\psi$ is a 9 component color-flavor spinor; $I=\delta^{\alpha\beta}\delta_{ij}$, 
$E$ and $Q$
are color-flavor matrices,
$E^{\alpha\beta}_{ij}=\delta^{\alpha\beta}(\varepsilon_{u}\delta_{i1}\delta_{j1}
+\varepsilon_{d}\delta_{i2}\delta_{j2}+\varepsilon_{s}\delta_{i3}\delta_{j3})$
and
$Q^{\alpha\gamma}_{ij}=3\Delta^{\gamma\alpha}_{ij}-\Delta^{\alpha\gamma}_{ij}$. 
${\it L}_{int}$ generates also the condensate  
$(1/G)Q^{\alpha\gamma}_{ij}\Delta^{\alpha\gamma}_{ij}$, equal to
\begin{equation}
-\frac{1}{G}2\left(b^2+10be+e^2+d^2-2(c^{2}-6cf+f^{2})\right)
\label{eq:a.2}\end{equation}
Diagonalizing the effective action matrix, Eq.~(\ref{eq:a.1}),
we obtain the free energy
as a sum of the quasiparticle energies and the condensate term.
Variations of the free energy with respect to the gap parameters
give the gap equations. 
Matrix $Q_{ij}^{\alpha\beta}$  
is block-diagonal in the color-flavor space,
that permits to diagonalize corresponding parts of the effective action separately.
Unitary transforming part of the effective action 
for the indices $(1,1),(2,2),(3,3)$, \cite{AlfordBergesRajagopal},
we obtain the 
$3\times 3$ part of
the condensate $Q_{ij}^{\alpha\beta}$
and the kinetic term $E^{\alpha\gamma}_{ij}$ 
\begin{equation}
\left(\begin{array}{ccc}
3b-e&0&0\\
0&b+5e&-\sqrt{2}(c-3f)\\
0&-\sqrt{2}(c-3f)&2d
\end{array}\right)\,,\;\;
\left(\begin{array}{ccc}
{\bar\varepsilon}&-\delta\varepsilon&0\\
-\delta\varepsilon&{\bar\varepsilon}&0\\
0&0&\varepsilon_{s}
\end{array}\right)
\label{eq:a.3}\end{equation}
correspondigly, where ${\bar\varepsilon}=\frac{1}{2}(\varepsilon_{u}+\varepsilon_{d})$
and
$\delta\varepsilon=\frac{1}{2}(\varepsilon_{u}-\varepsilon_{d})$.
The $2\times 2$ parts of the $Q_{ij}^{\alpha\beta}$ are off-diagonal
containing matrix elements $(3b-e)$ for $(1,2),(2,1)$ indices, and $(3c-f)$
for $(1,3),(3,1)$ and $(2,3),(3,2)$ indices.

{\it Mixed BCS-BP three flavor state.}
When $|\varepsilon_{u}-\varepsilon_{d}|\ll|\varepsilon_{u}-\varepsilon_{s}|\sim
|\varepsilon_{d}-\varepsilon_{s}|$, i.e.  
$\delta\varepsilon\sim 0$, 
system of the eigenvalue equations for the $3\times 3$-block, Eq.~(\ref{eq:a.1}), 
decouples,
producing two quasiparticle energies with the gap $(3b-e)$ and four quasiparticle energies 
satisfying
\begin{eqnarray}
\lambda^{4}&-&\lambda^{2}\left[{\bar\varepsilon}^2+\varepsilon_{s}^{2}+(b+5e)^2
+4(c-3f)^2+4d^2\right] \\
&+&\left({\bar\varepsilon}^2+(b+5e)^2\right)\left(\varepsilon_{s}^2+4d^2\right)
+4(c-3f)^2\left({\bar\varepsilon}\varepsilon_{s}-2d(b+5e)^{\phantom{1}}
\hspace{-0.2cm}\right)
+4(c-3f)^{4}=0 \nonumber
\label{eq:a.4}\end{eqnarray}    
where the eigenvalue energy is $\lambda+p_{0}$. 
All quasiparticle energies with their degeneracies are
\begin{eqnarray}
\begin{array}{lr}
p_{0}\pm\sqrt{\varepsilon_{u}^2+(3b-e)^2}&(1)\\
p_{0}\pm\frac{1}{\sqrt{2}}
\sqrt{\varepsilon_{u}^2
+\varepsilon_{s}^2+(b+5e)^2+4(c-3f)^2+4d^2\pm\sqrt{D}}&
(2)\\
p_{0}\pm\sqrt{\varepsilon_{u}^2+(3b-e)^2}&(2)\\
p_{0}+\frac{1}{2}(\varepsilon_{u}-\varepsilon_{s})\pm
\sqrt{(\frac{1}{2}(\varepsilon_{u}+\varepsilon_{s}))^2+(3c-f)^{2}}&(2)\\
p_{0}-\frac{1}{2}(\varepsilon_{u}-\varepsilon_{s})\pm
\sqrt{(\frac{1}{2}(\varepsilon_{u}+\varepsilon_{s}))^2+(3c-f)^{2}}&(2)
\end{array}
\label{eq:a.5}\end{eqnarray}
where
\begin{eqnarray}
D &=& \left[\varepsilon_{u}^{2}+\varepsilon_{s}^{2}+(b+5e)^{2}
+4(c-3f)^{2}+4d^{2}\right]^{2}
\\
&-& 4\left[(\varepsilon_{u}^{2}+(b+5e)^{2})(\varepsilon_{s}^{2}+4d^{2})
+4(c-3f)^{2}(\varepsilon_{u}\varepsilon_{s}-2d(b+5e))+4(c-3f)^{4}
\right]\nonumber
\label{eq:a.6}\end{eqnarray}
We define $S_{\pm}=\frac{1}{\sqrt{2}}\sqrt{\varepsilon_{u}^2
+\varepsilon_{s}^2+(b+5e)^2+4(c-3f)^2+4d^2\pm \sqrt{D}}$. 
There are $(9)\times 2$ eigenvalues.  
In the CFL limit ($b=c$, $e=f$, $d=b+e$, $\varepsilon_{u}=\varepsilon_{s}$),
these eigenvalues reduce to $(8)$ modes with the gap $\Delta_{8}=3b-e$ and 
$(1)$ mode with the gap $\Delta_{1}=8e$, manifesting the 
$SU(3)_{V}$ symmetry where $V=color+flavor$. Strange quark mass 
breaks the $SU(3)_{V}$ 
and $8$ modes split into isomultiplets, $8=3+2+2+1$.
In the $\bar{3}$ channel these quasiparticle energies were obtained in \cite{SchaeferWilczek}.
Varying the free energy, given by the sum of quasiparticle energies
Eq.~(\ref{eq:a.5}) and the condensate Eq.~(\ref{eq:a.2}), 
with respect to the $5$ gap parameters,
$b, e, c, f, d$:
$\frac{\partial\langle H\rangle}{\partial b}=0$ etc., 
we obtain $5$ gap equations. Since 
$S_{\pm}$ is a function of $(b+5e)$ and $(c-3f)$, 
$5\frac{\partial S}{\partial b} 
=\frac{\partial S}{\partial e}$
and $3\frac{\partial S}{\partial c}=-\frac{\partial S}{\partial f}$.
Combining, $5\frac{\partial\langle H\rangle}{\partial b}
-\frac{\partial\langle H\rangle}{\partial e}=0$
and $3\frac{\partial\langle H\rangle}{\partial c}
+\frac{\partial\langle H\rangle}{\partial f}=0$, we obtain
in the $\bar{3}$ channel ($e=-b$, $f=-c$, $d=0$),
\begin{eqnarray}
&& \frac{1}{G}b+2\int\frac{d^{3}{\bf p}}{(2\pi)^{3}}
\frac{b}{\sqrt{\varepsilon_{u}^{2}+(4b)^{2}}}=0
\nonumber\\
&& \frac{1}{G}c+2\int_{Q}\frac{d^{3}{\bf p}}{(2\pi)^{3}}\frac{c}
{\sqrt{(\frac{1}{2}(\varepsilon_{u}+\varepsilon_{s}))^{2}+(4c)^{2}}}=0
\label{eq:a.7}\end{eqnarray}
describing the BCS gap $\Delta_{ud}=4b$ and
the breached pairing gap $\Delta_{us}\sim \Delta_ {ds} =4c$,
where $Q=\{\,|\frac{1}{2}(\varepsilon_{u}+\varepsilon_{s})|
>\sqrt{(\frac{1}{2}(\varepsilon_{u}-\varepsilon_{s}))^{2}-(4c)^{2}}\,\}$ 
is the BP momentum integration area.
Eq.~(\ref{eq:a.7}) contains the two-flavor gap equations,
Eq.~(\ref{eq:0.3}), with $G=g/4$.

{\it BP two flavor state.}
When
$\delta \varepsilon\neq 0$ and $\delta \varepsilon\ll |{\bar\varepsilon}-\varepsilon_{s}|$,
the gap parameters $c=f=d=0$,
and the eigenvalue equation for the $3\times 3$ block reads
 \begin{eqnarray}
\lambda^{4} &-& 2\lambda^{2}\left({\bar\varepsilon}^{2}
+\frac{1}{2}((3b-e)^{2}+(b+5e)^{2})
+\delta\varepsilon^{2}\right)
+ \left({\bar\varepsilon}^{2}+\frac{1}{2}((3b-e)^{2}+(b+5e)^{2})
-\delta\varepsilon^{2}\right)^{2}
\nonumber\\
&+&\left[\delta\varepsilon^{2}\left((3b-e)+(b+5e)\right)^{2}
-\frac{1}{4}\left((3b-e)^{2}-(b+5e)^{2}\right)^{2}\right]=0
\label{eq:a.8}\end{eqnarray}   
Solving this equation we omit
terms in the square bracket, since they
do not contribute to the gap equation in the ${\bar 3}$ channel ($e=-b$).
The quasiparticle energies
are  
\begin{eqnarray}
\begin{array}{lr}
p_{0}+\delta\varepsilon\pm\sqrt{{\bar\varepsilon}^{2}
+\frac{1}{2}((3b-e)^{2}+(b+5e)^{2})}& (1)\\
p_{0}-\delta\varepsilon\pm\sqrt{{\bar\varepsilon}^{2}
+\frac{1}{2}((3b-e)^{2}+(b+5e)^{2})}& (1)\\
p_{0}+\delta\varepsilon\pm\sqrt{{\bar\varepsilon}^{2}+(3b-e)^{2}}& (1)\\
p_{0}-\delta\varepsilon\pm\sqrt{{\bar\varepsilon}^{2}+(3b-e)^{2}}& (1)
\end{array}
\label{eq:a.9}\end{eqnarray}
which are $(4)\times 2$ modes.
Condensate is 
$\frac{1}{G}2\left(b^{2}+10be+e^{2}\right)$.
Combining, $3\frac{\partial\langle H\rangle}{\partial b}
+\frac{\partial\langle H\rangle}{\partial e}=0$,
and taking the ${\bar 3}$ channel,
we obtain
\begin{eqnarray}
\frac{1}{G}b+2\int_{Q}\frac{d^{3}{\bf p}}{(2\pi)^{3}}\frac{b}
{\sqrt{{\bar\varepsilon}^{2}+(4b)^{2}}}=0
\label{eq:a.10}\end{eqnarray}
describing the breached pairing gap $\Delta_{ud}=4b$,
where $Q=\{\,|\bar\varepsilon|
>\sqrt{\delta\varepsilon^{2}-(4b)^{2}}\,\}$ 
is the BP momentum integration area.

\end{document}